\documentclass[fleqn,usenatbib]{mnras}

\usepackage{newtxtext,newtxmath}

\usepackage[T1]{fontenc}
\usepackage{ae,aecompl}

\usepackage{graphicx}	
\usepackage{amsmath}	
\usepackage{amssymb}	
\usepackage{soul}
\usepackage{caption}
\usepackage{hyperref}
\usepackage{dblfloatfix}
\usepackage{xcolor}

\title[]{8 in 10 Stars in the Milky Way Bulge Experience Stellar Encounters Within 1000 AU in a Gigayear}

\author[McTier, Kipping, \& Johnston]{
Moiya A.S. McTier,$^{1}$\thanks{E-mail: mmctier@astro.columbia.edu}
David M. Kipping,$^{1,2}$ Kathryn Johnston$^{1,2}$
\\
$^{1}$Department of Astronomy, Columbia University, 550 W 120th St., New York, NY 10027\\$^{2}$Center for Computational Astrophysics, Flatiron Institute, 162 5th Ave, New York, NY 10010\\
}

\date{Accepted XXX. Received YYY; in original form ZZZ}

\pubyear{2019}

\begin{document}
\label{firstpage}
\pagerange{\pageref{firstpage}--\pageref{lastpage}}
\maketitle

\begin{abstract}
The Galactic bulge is a tumultuous dense region of space, packed with stars separated by far smaller distances than those in the Solar neighborhood. A quantification of the frequency and proximity of close stellar encounters in this environment dictates the exchange of material, disruption of planetary orbits, and threat of sterilizing energetic events. We present estimated encounter rates for stars in the Milky Way bulge found using a combination of numerical and analytical methods. By integrating the orbits of bulge stars with varying orbital energy and angular momentum to find their positions over time, we were able to estimate how many close stellar encounters the stars should experience as a function of orbit shape. We determined that $\sim$80\% of bulge stars have encounters within 1000 AU and that half of bulge stars will have >35 such encounters, both over a gigayear. Our work has interesting implications for the long-term survivability of planets in the Galactic bulge.
\end{abstract}




\section{Introduction}
Most, though not all, stars form in clusters \citep{bressert:2010}. The vast majority of these clusters (>90\%) contain only $10^{2-3}$ stars per cubic parsec and are therefore too sparsely populated to withstand dynamical stresses over many orbital periods. As a result, these open clusters evaporate within $\sim10^8$ years or so \citep{lada:2003}. Our Sun and nearby stars were likely born in such sparse open cluster environments. 

Globular clusters, however, are dense enough ($10^4$ stars per cubic parsec) to hold together under those dynamical stresses, and are particularly interesting environments to study for a couple of reasons. First, the relative certainty of stellar ages derived from cluster ages \citep{soderblom:2010} makes it possible to study both stars and planets at known evolutionary snapshots. Second, the long lifespans and high stellar density of globular clusters increase the probability of close stellar encounters occurring.

Several groups have studied the consequences of close stellar encounters, which can affect sterilization of planetary systems, the exchange of material between systems, planet formation, and planet survivability. Most of these studies seem to focus on the latter relationship between stellar encounters and planet survivability. 

For example, close encounters can strip planets away from their hosts or destabilize their orbits in the long-term \citep{spurzem:2009, malmberg:2011, zheng:2015, yang:2015}. \citet{portegies:2015} determined the orbital parameters for which a planet may be considered safe from perturbations from sources outside its host system. \citet{li:2019} simulated stellar encounters for Solar System analogs in open clusters and found that 25\% of outer planets in a system like ours can be lost or captured by the fly-by star. \citet{elteren:2019} and \citet{cai:2019} find that $\sim$14\% of planets in a dense stellar cluster will be lost from their stars within $\sim 10^7$ years of their formation, but the majority of these orphaned worlds have initial semimajor axes >20 AU, which is consistent with previous findings \citep{parker:2012}.

Despite the wealth of research on planet survivability in dense environments, less than 1\% of confirmed exoplanets orbit stars found in clusters \citep{cai:2017}, which does imply a difference between planetary systems in and out of dense stellar environments. \citet{meibom:2013}, however, found that there is \textit{not} a dearth of planets in dense stellar cluster environments, though most of the planets are either free-floating, unstable in their orbits, or have short periods.

\begin{figure*}
\includegraphics[width=\textwidth]{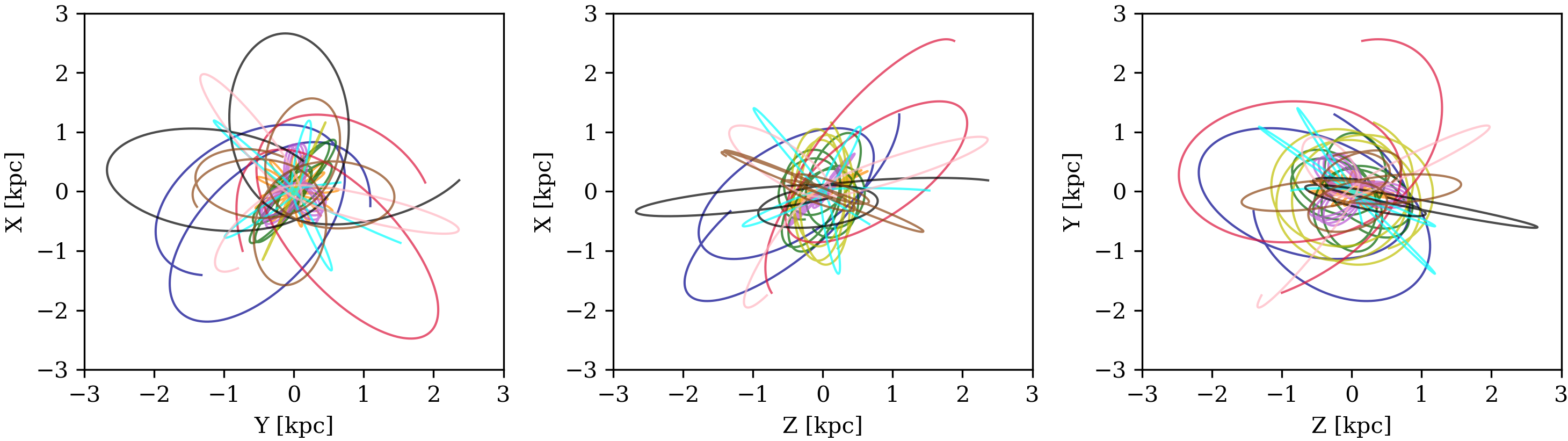}
\caption{Integrated orbits of 10 randomly selected stars over 10$^8$ years. Note that stars aren't moving on circular orbits, but are instead on eccentric rosette orbits characteristic of the Galactic bulge.}
\label{fig:sample_orbits}
\end{figure*}

This led us to wonder how common stellar encounters might be in the Milky Way bulge, which has a similar stellar density to the densest globular clusters \citep{valenti:2016}, but a much higher velocity dispersion ($\sim120$ km/s for the bulge \citep{valenti:2018} compared to $\sim10$ km/s for globular clusters \citep{baumgardt:2019}).

Most of the previous studies focused on close stellar encounters and planet survival specifically in dense cluster environments. That's not surprising given that the vast majority of confirmed exoplanets orbit stars less than 100pc away from the Sun (according to the NASA Exoplanet Archive \citep{akeson:2013}) due to instrument and survey limitations. A small fraction of confirmed exoplanets, however, sit far outside of the Solar neighborhood. \citet{sahu:2006} proposed 16 transiting planet candidates in the Milky Way bulge and \citet{batista:2013} reported the first exoplanet to be identified as belonging to the bulge with high confidence. These planets are more than 8kpc from the Sun.

Little is known about the fate of planetary systems in the Galactic bulge. The planet occurrence rate has been well constrained in the Solar neighborhood \citep{hsu:2019, hardegree:2019}, for example, but it's impossible to glean a reliable planet occurrence rate for the bulge based on the detection of only a few confirmed planets. Knowing the stellar encounter rate for the Galactic bulge can help us constrain our expectations for its planet occurrence rate, as well as better understand sterilization, material exchange, and a host of other phenomena in this mysterious environment.

\citet{jimenez:2013} studied the effects of close stellar encounters on planet survival in different galactic environments, including the bulge. Using a density profile and velocity dispersion for the region, they estimated the frequency and distance of stellar fly-bys, and then simulated planetary disk responses to such encounters. They determined that stars in the bulge can experience up to $\sim$400 encounters within 200 AU in 4.5 Gyr. 

Our goal in this work was see how a bulge stellar encounter rate determined using different computational tools compares to the rate found by \citet{jimenez:2013}. We also wanted to add new analysis of how encounter rates vary by stellar orbit shape as defined by the star's energy and angular momentum, could potentially be determined using near future observations. To that end, we simulated the orbits of stars with different energies and angular momenta to find their positions over time and analytically estimated their encounter rates.


\section{Methods}\label{sec:calc}
In order to estimate the encounter rate for stars in the Milky Way bulge, we employed a semi-analytic method that combined numerical integration of stellar orbits and analytical estimates of stellar number density. Here, we describe the steps of that process in more detail.

\subsection{Simulation Setup}\label{sec:init}
The initial positions and velocities for our particles were generated using a code that builds a density profile according to the Hernquist potential \citep{hernquist:1990}, which closely resembles the density distribution of the bulge. 

We set the scale length of the bulge to $a=0.31$ kpc \citep{li:2016, hernquist:1990} and the stellar mass of the bulge to $M_b = 2\times10^{10}M_{\odot}$ \citep{valenti:2016}.

The orbit simulation is done using $\mathtt{gala}$'s \citep{price-whelan:2017} Leapfrog integrator in the Hernquist potential, which uses each star's position and velocity at each timestep to calculate where that star would be after time $\Delta$t assuming a stable Hernquist gravitational potential. The equilibrium nature of the distribution ensures that the density profile of the set of particles remains constant. 


In order to save computational time and because of the long relaxation timescale (the amount of time it takes one object in the system to be significantly perturbed by close encounters with other objects in the system) of the bulge \citep{binney:1988}, we chose to simulate massless non-interacting particles. After all, we're interested in the way that close stellar encounters perturb the orbits of \textit{planets}, not the orbits of their host stars.

Stellar orbits in the Galactic bulge are morphologically and kinematically different from those in the disk. While stars in the disk move on more or less circular orbits (with some epicyclic motion), bulge stars react to the spherical mass distribution by moving on more eccentric orbits that can outline rosette shapes over time. Relatedly, disk star velocities approximately follow Kepler's third law because the disk's structure is supported by its motion. The bulge's structure, however, is supported largely by pressure, so bulge star velocities are random and therefore not as simple to predict. For context, Figure~\ref{fig:sample_orbits} shows the orbits of 10 randomly selected bulge stars integrated over 10$^8$ years.

\subsection{Numeric Estimation of Encounter Rate}
Because the orbit integrator tracks the position of every particle at each time step, it's possible to numerically determine how many stars have encounters within a designated distance by counting the number of neighboring particles in the system  within a designated distance at each timestep. We used $\mathtt{scipy.spatial}$'s $k$-D Tree \citep{manee:1999}, a type of binary space partitioning tree that makes it easy to quickly search through multidimensional space. 

If given a $k$-dimensional data array, the function performs a series of binary splits where it chooses an axis and separates the data into two groups along that axis. After many such splits, each data point is then assigned to a \textit{node} -- one of the smaller divided regions -- which makes it easier to search through the space. $\mathtt{scipy}$'s $k$-D Tree function implements the algorithm described in \citet{manee:1999} to find the nearest neighbor to the points we specify. It's not a perfect method as the function only searches for neighbors that reside in the same node as the specified point -- it's possible that the real nearest neighbor is in an adjacent node and won't be considered -- but it is fast and provides a good approximation for the nearest neighbor distances we wanted to find.

\subsection{Semi-Analytic Estimation of Encounter Rate}
Numerically simulating encounter rates for all $10^{10}$ stars in the Galactic bulge \citep{licquia:2015, picaud:2004} is an inefficient use of computer time. Instead, we wanted to see if a semi-analytic estimation would match the numerically-determined encounter rates for simulations with lower stellar number densities.

The semi-analytic calculation is done by integrating stellar orbits to get the stars' galactocentric positions over time, which also determines the change in stellar number density (see Eqn~\ref{eqn:rho}) and encounter velocity over time. With the stellar number density and encounter velocity, we estimated the encounter rate using

\begin{equation}\label{eqn:col}
Z(r,h,\Delta t) = \sum \pi h^2 v_{\textrm{enc}}(r) \rho(r) \Delta t
\end{equation}

\noindent where $\rho(r)$ is the stellar number density as a function of galactocentric radius, given by 

\begin{equation}\label{eqn:rho}
\rho(r) = \frac{N}{2\pi}\frac{a}{r}\frac{1}{(r+a)^3}
\end{equation}

\noindent according to the Hernquist profile \citep{hernquist:1990}. The density depends on the number of stars $N$, galactocentric radius $r$, and scale radius $a= 0.31$ kpc as noted in \S~\ref{sec:init}. $Z(r,h,t)$ is the number of encounters within a distance $h$ over a timescale $\Delta t$. We take $v_{\textrm{enc}}$ to depend on both the instantaneous velocity of the star ($v$) and the velocity distribution ($\sigma_v$) of the bulge at the star's position, which are calculated at each step of the integration. 

\begin{figure}
\includegraphics[width=3.2in]{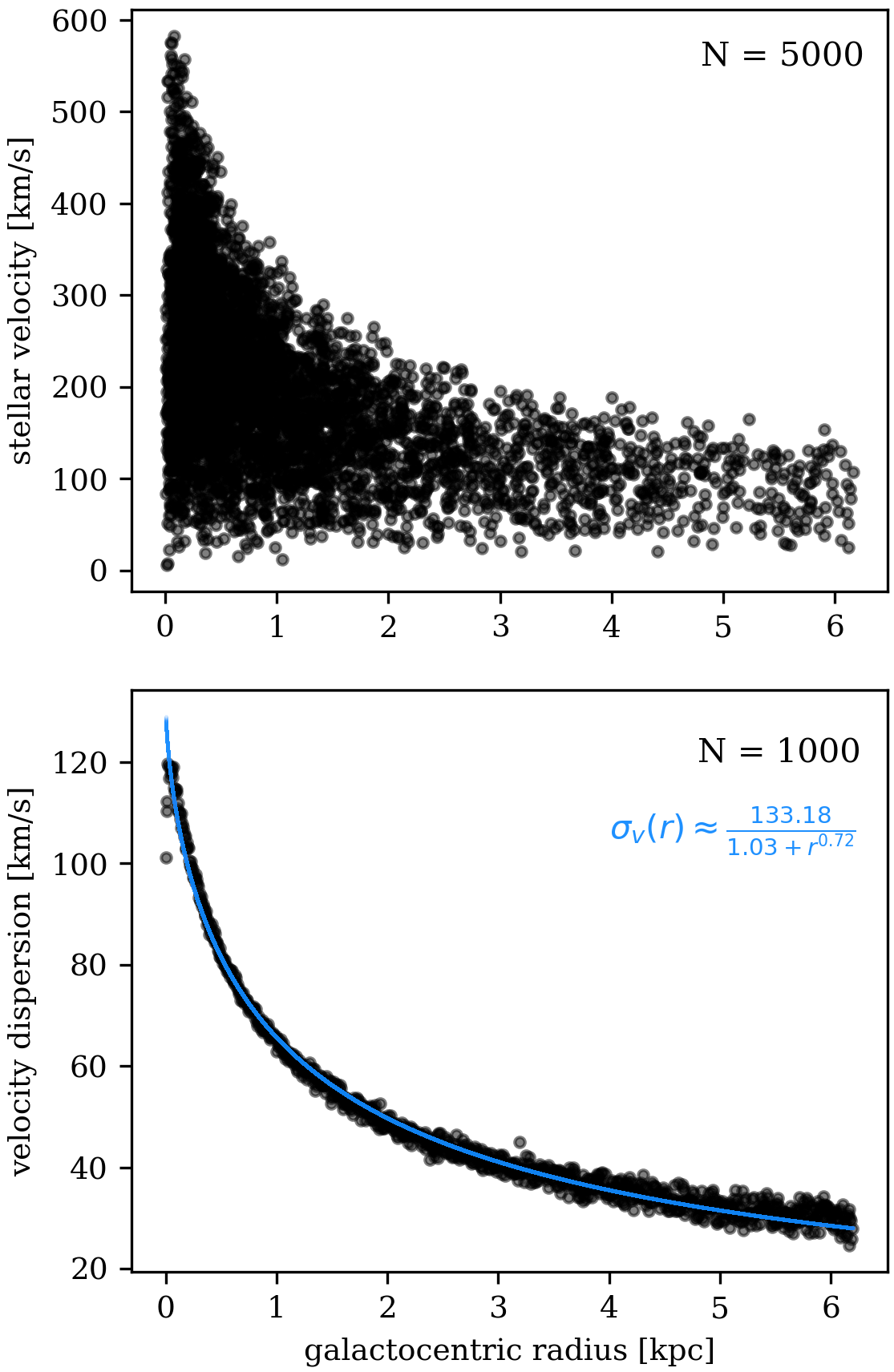}
\caption{Top: total velocities ($\sqrt{V_x^2 + V_y^2 + V_z^2}$) at different galactocentric radii. Bottom: velocity dispersion as a function of galactocentric radii (black) and dispersions calculated using Eqn~\ref{eqn:vel_disp} (blue)}
\label{fig:vel_disp}
\end{figure}

We found the velocity dispersion profile for the bulge by separating the initial velocities (see \S~\ref{sec:init}) into bins and calculating the standard deviation within each bin. Using $\mathtt{scipy}$'s Curve Fit function, we found that the velocity dispersion profile can be described (as evident from Figure~\ref{fig:vel_disp}) by 

\begin{equation}\label{eqn:vel_disp}
\sigma_v(r) \approx \frac{133.18}{1.03+r^{0.72}}.
\end{equation}

The exact analytical expression for the velocity dispersion of the bulge is given in Eqn. 10 of \citet{hernquist:1990}. Note that our model does not take into account the influence of the disk and dark matter halo in which the bulge is embedded. These would raise the velocity dispersion at higher galactocentric radii, increasing the encounter rates, so our results in these regions should be considered a lower limit. 

Assuming isotropic, Gaussian velocity distributions with dispersion, the characteristic encounter velocity is then given by 

\begin{equation}\label{eqn:venc}
    v_{\textrm{enc}}(r) = \sqrt{v^2 +\sigma_{v}(r)^2}.
    \end{equation} 

Note that this expression differs from that used to describe the average relative speed between two stars selected at random from a Gaussian distribution (where $v_{\textrm{enc}} = \sqrt(2) \sigma_{v}$ \citep{mihos:2003}) because we are considering the encounter rates of individual stars along specific orbits.

\subsection{Numerical versus Analytical Estimates}
To see if the semi-analytic estimation of encounter rates matched the numerical estimates, we followed the orbits of 1000 randomly selected stars in our sample and determined the encounter rates both numerically and semi-analytically for each so that we could directly compare them.

The simulation itself integrated the orbits of $10^6$ stars, each for one radial period (determined using Eqn.~\ref{eqn:rad_per}) divided evenly into 100 steps. For this test, we set $h = 10^7$ AU to ensure that there would actually be encounters in such a low-density environment.  

\begin{figure}
\includegraphics[width=3.2in]{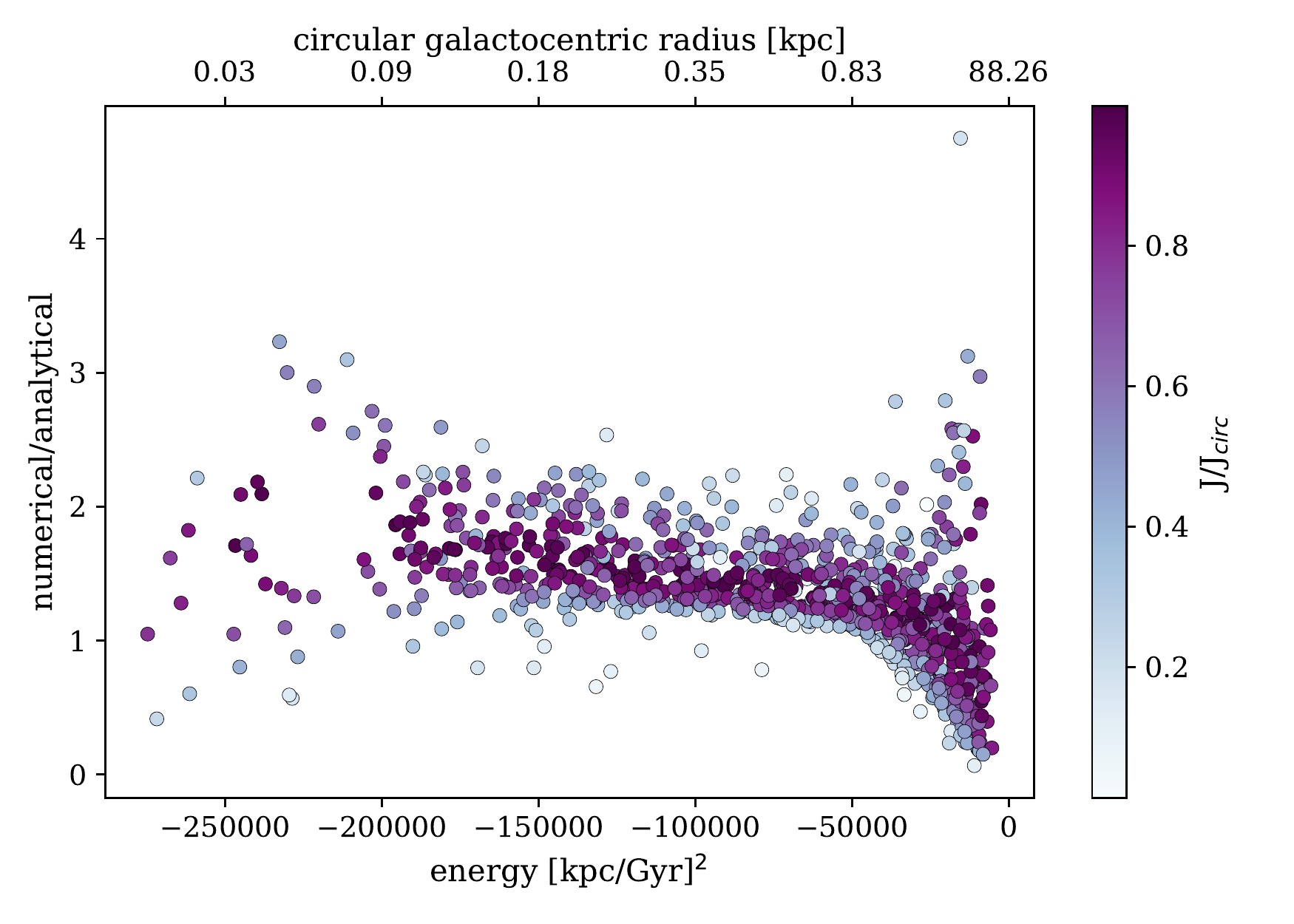}
\caption{Ratios of encounter rates determined numerically to those determined semi-analytically as a function of orbital energy for 1000 randomly selected stars in our sample. Points are colored according to $J/J_{\textrm{circ}}$, a metric that quantifies how circular the star's orbit is.}
\label{fig:ratios}
\end{figure}

Figure~\ref{fig:ratios} shows the ratios of encounter rates determined numerically to those determined semi-analytically.

\begin{figure*}
\includegraphics[width=\textwidth]{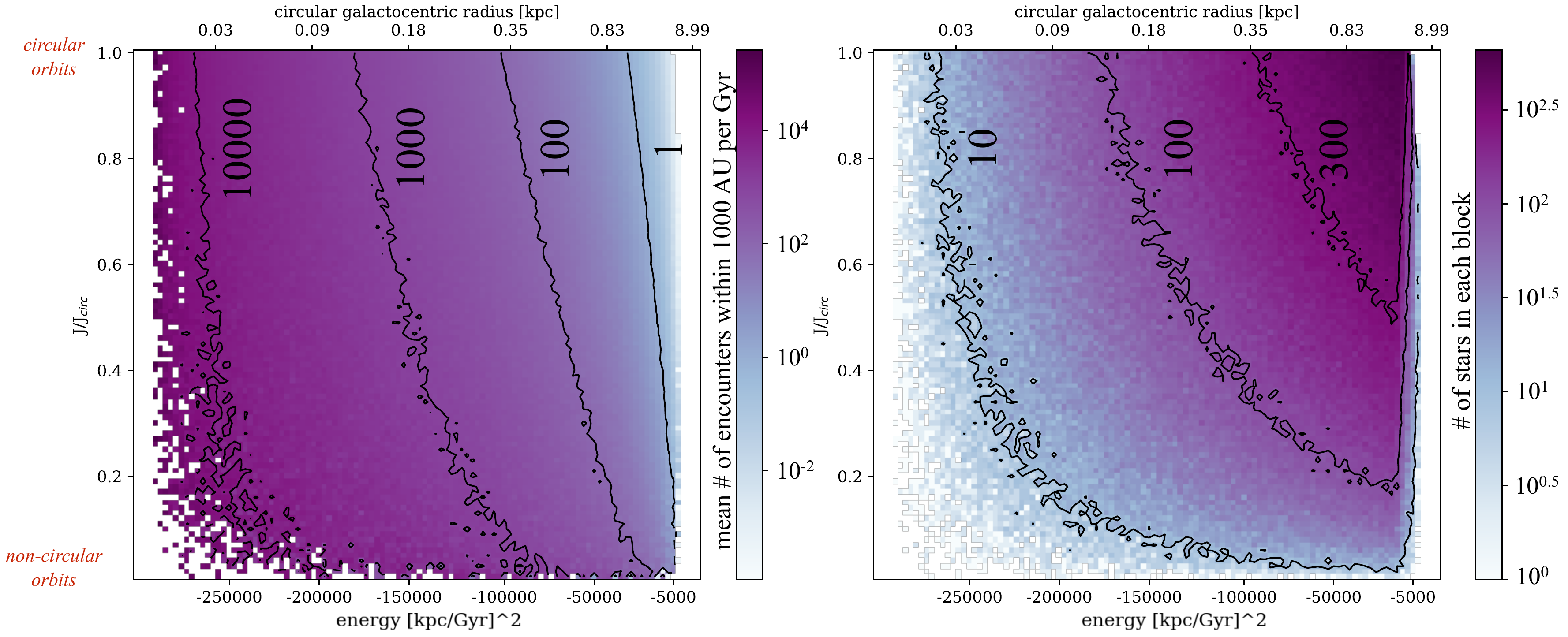}
\caption{Left: mean encounter rates for encounters within 1000 AU as a function of energy and angular momentum. The rates are determined by numerically integrating $10^6$ orbits and analytcally estimating the number of encounters with Eqn~\ref{eqn:col} at each point, setting $N=10^{10}$ in Eqn~\ref{eqn:rho}. Right: number of stars that in each block of the $E$-$J/J_{\textrm{circ}}$ grid in our $10^6$ star sample.}
\label{fig:encounters_grid}
\end{figure*}

The median ratio between numerical and semi-analytical encounter rates is $\sim$1.3. We attribute this offset to the fact that the encounter velocity we use (see Eqn~\ref{eqn:venc}) doesn't precisely account for the velocity of the other star involved in the encounter. Also, the numerical method yields an integer number of encounters while the analytical method yields a floating decimal; the ratio of the two estimates will not be exactly 1. 

The numerical and semi-analytical estimates are all within an order of magnitude of each other and are within a factor of 2 for $\sim 90$\% of the trials. This scatter is likely due to the fact that our semi-analytic method depends on the timestep used during the integration in that shorter timesteps yield more accurate estimates. Computational constraints keep us from using very small $\Delta t$s. 

We ran tests to confirm that the ratio of numerical-to-analytical estimates stays the same regardless of stellar number density ($N$) and encounter distance ($h$) used, so we are confident that that our ratio holds true for the bulge, which has $\sim 10^{10}$ stars \citep{licquia:2015, picaud:2004}. 

Figure~\ref{fig:ratios} also shows that the ratio of numerical to analytical estimates is irrespective of the orbit's eccentricity and only mildly dependent on binding energy for orbits within a few kpc of the bulge.

Overall, we conclude that the semi-analytical estimate of encounter rates sufficiently matches the numerical estimate, and we adopt the analytical estimate for the rest of the paper without using any sort of scale factor.

\subsection{Defining Orbits by Energy and Angular Momentum}
One of our goals was to determine the encounter rate of stars in the Milky Way bulge as a function of orbit shape. Because orbital energy ($E$) and angular momentum ($J$) are conserved throughout a star's orbit, they can be used together to define an orbit's shape. We calculated $E$ and $J$ from the initial positions and velocities of $10^6$ stars using

\begin{equation}\label{eqn:J}
J = r\times v
\end{equation}

\noindent and

\begin{equation}\label{eqn:E}
E = \frac{1}{2}v^2 - \frac{GM}{r+a}
\end{equation}

\noindent where $M$ is the mass of the system (the Galactic bulge, in this case), and not the mass of the star.

We anticipated that the encounter rate for stars on circular orbits would have a stronger relation to galactocentric radius than the rate for stars on more irregular orbits. To account for this, we normalized each star's $J$ by the angular momentum of a star on a circular orbit with the same energy. To find that $J_{\textrm{circ}}$, we first had to find the radius of a circular orbit with a certain energy, given by

\begin{equation}\label{eqn:r_circ}
r_{\textrm{circ}} = \textrm{max}\left (\pm\frac{\sqrt{GM(GM - 8aE)} \mp 4aE \mp GM}{4E}\right )
\end{equation}

\noindent which could then be substituted into the expanded equation for angular momentum

\begin{equation}\label{eqn:J_circ}
J_{\textrm{circ}} = r_{\textrm{circ}}\sqrt{\frac{GMr_{\textrm{circ}}}{(r_{\textrm{circ}}+a)^2}}
\end{equation}

We separated our $10^6$ stars into a grid according to their energy $E$ and angular momentum normalized by that of a circular orbit with the same energy $J/J_{\textrm{circ}}$. We then integrated each star for one radial period (the amount of time it takes to go from the star's pericenter to apocenter and back again) with 10,000 steps.

For a given $E$ and $J$, the pericenter and apocenter are the extrema of a star's galactocentric position and can be found numerically. We found them using $\mathtt{scipy.optimize}$'s Bisect function to get the roots of Eqn~\ref{eqn:roots}.

\begin{equation}\label{eqn:roots}
\dot{r}^2 = 2(E+\frac{GM}{r+a}) - (J/r)^2
\end{equation}

We then found the radial periods by integrating 

\begin{equation}\label{eqn:rad_per}
T = 2\int_{\textrm{peri}}^{\textrm{apo}} \frac{1}{\sqrt{2(E+\frac{GM}{r+a}) - J^2/r^2}}dr
\end{equation}

\noindent (\citet{binney:1988}, but see \citet{bovy:2017} for a summary) from the pericenter to apocenter.

We integrated each star's orbit over its radial period and used its time-varying position to estimated the encounter rates semi-analytically. We describe our results in the next section.

\section{Results \& Comparison}\label{sec:results}
\begin{figure}
\includegraphics[width=3.2in]{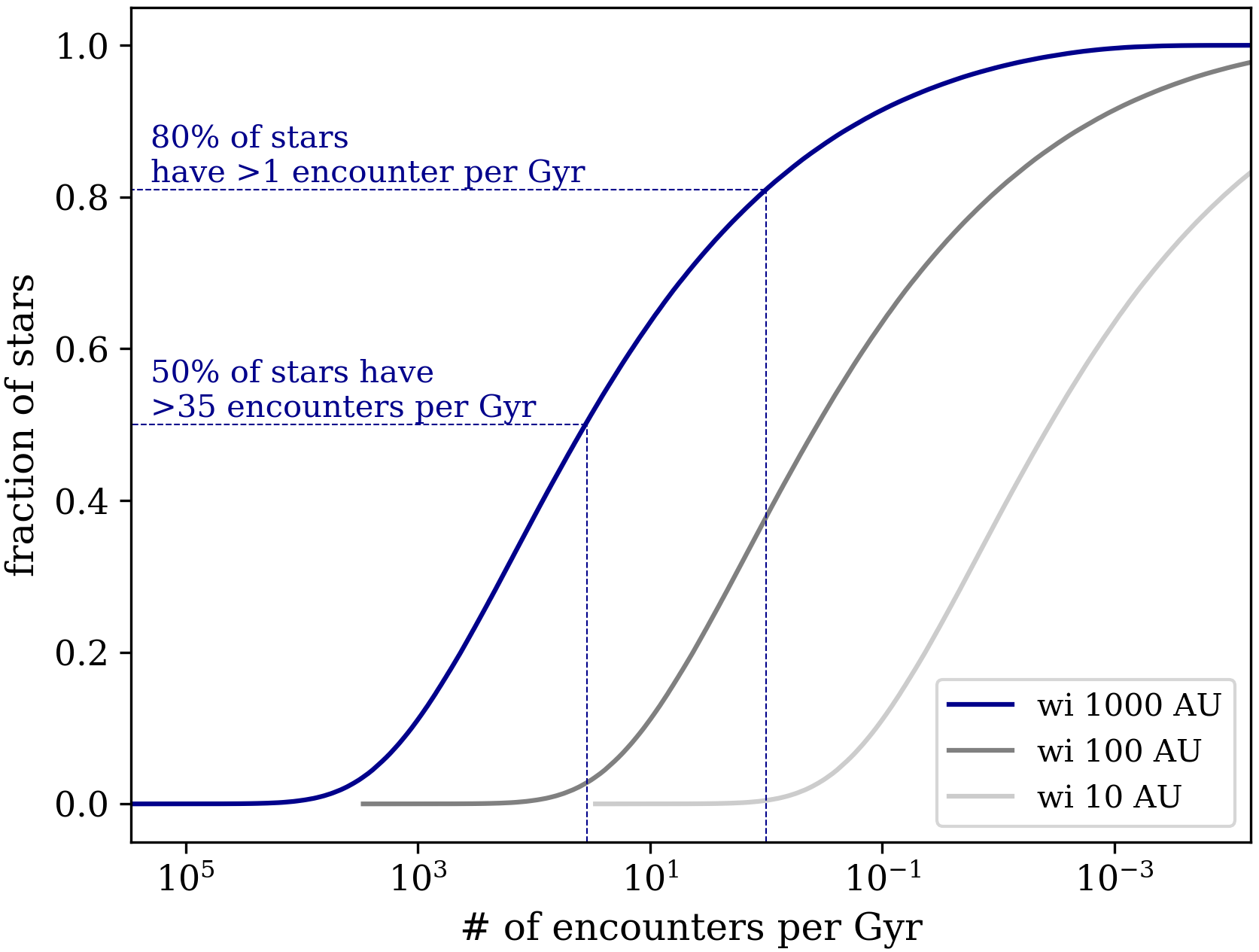}
\caption{Cumulative distribution function of encounter rates within 1000 AU, 100 AU, and 10 AU. The rates are determined by numerically integrating $10^6$ orbits and analytcally estimating the number of encounters with Eqn~\ref{eqn:col} at each point, setting $N=10^{10}$ in Eqn~\ref{eqn:rho}.}
\label{fig:encounters_cdf}
\end{figure}

After confirming that a semi-analytical estimate matched a numerical estimate to our satisfaction, we once again simulated the orbits of $10^6$ stars. This second time, we integrated each star for a total of one radial period evenly divided into 10000 steps. At each step, we used Eqns ~\ref{eqn:col} and ~\ref{eqn:rho} to determine the encounter rates and artificially increased the stellar number density by setting $N = 10^{10}$. We set $h = 1000$ AU as a fiducial encounter distance because it's the typical threshold used to define a close stellar encounter \citep{adams:2001, adams:2006} as most interactions within this range will perturb or disrupt planetary orbits or formation processes.

We converted our number of encounters per orbit to encounters per gigayear. Each block in the $E$-$J/J_{\textrm{circ}}$ had several stars (number shown in the right-hand panel of Figure~\ref{fig:encounters_grid}), and we took the mean encounter rate for each block. The left-hand panel of Figure~\ref{fig:encounters_grid} shows the mean encounter rate for encounters within 1000 AU as a function of energy and angular momentum. 

You can see that stars experience fewer encounters as their orbits move further away from the galactic center. This trend is even stronger for stars that move on more circular orbits, as the stellar density of their environment doesn't change over time. As noted above, we expect to underestimate the encounter rate in these regions as we do not account for gravitational influence from the disk or dark matter halo.

Figure~\ref{fig:encounters_grid} doesn't represent all of the stars in our sample, however, because each value in the grid is the average of many encounter rates. Figure~\ref{fig:encounters_cdf} shows the cumulative distribution function (CDF) of all encounter rates for encounters within 1000 AU.

You can see in Figure~\ref{fig:encounters_cdf} that 50\% of stars in the Milky Way bulge will experience more than 35 encounters within 1000 AU in just 1 Gyr, $\sim$80\% of stars have more than one of these encounters in a Gyr, and $\sim$35\% of stars have more than one encounter within 100 AU in a Gyr. We expect less than 1 star in 5000 to have encounters within 10 AU.

This result can be compared to earlier work. \citet{jimenez:2013} determined that the number of stellar encounters in the Milky Way bulge should peak around 100 pc from the Galactic center at about 350 encounters within 200 AU over 4.5 Gyr. Our encounter rate also peaks around 100 pc from the Galactic center, and scaling our encounter estimates to similar circumstances yields a rate of $\sim$360 encounters within 200 AU over 4.5 Gyr. It's important to note that \citet{jimenez:2013} assumed that their stars moved on circular orbits and used the velocity dispersion (fit from sparse observations taken in 2002) as the typical encounter velocity, whereas we account for the complicated dynamics of the bulge and take the encounter velocity to be the instantaneous velocities of the stars as found by our simulation.  

\begin{figure*}
\includegraphics[width=\textwidth]{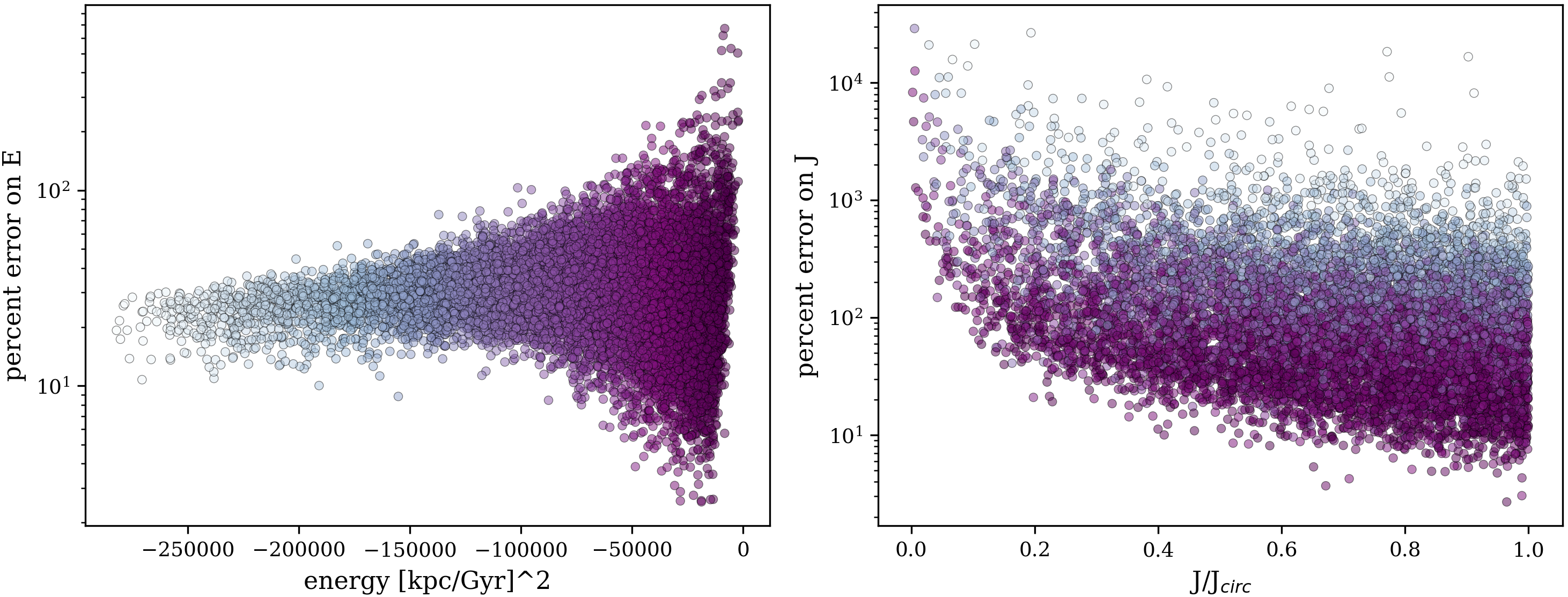}
\caption{Percent errors on energy and angular momentum, colored by energy so you can match stars across the two panels.}
\label{fig:E_J_err}
\end{figure*}

Though we used 1000 AU as a fiducial encounter threshold, our semi-analytic method of estimating encounter rates makes it possible to easily scale to other desired encounter distance thresholds. For example, dynamical studies indicate that stellar clusters with only $10^2$$^-$$^3$ members per cubic parsec can expect $\sim$10\% of the group to experience encounters within 100 AU before they dissipate \citep{li:2019,malmberg:2011}. To get rates of encounters within 100 AU, you can simply scale our reported values down by a factor of 100.

\section{Summary \& Discussion}\label{sec:summary}
We started this work because we saw the abundance of literature exploring the frequency and consequences of close stellar encounters in dense star clusters and wanted to know if the findings held for the Milky Way bulge, a region of the galaxy with similar density to but much greater velocity dispersion than the densest star clusters \citep{valenti:2016, valenti:2018, baumgardt:2019}. We have estimated the number of close stellar encounters that can be expected in the Milky Way bulge using a combination of numerical integration \citep{price-whelan:2017} and analytical methods. We found that $\sim$90\% of bulge stars should have an encounter within 1000 AU within a Hubble time. That number scales down for closer encounters.

This number may seem shockingly high, but it makes sense when you consider that the Sun is expected to have an encounter with the K dwarf GI 710 within 1600 AU in the next 1.3 Myr \citep{bailer-jones:2018}. If the Sun -- sitting at 8.3 kpc from the Galactic center \citep{gillessen:2009} in a region of the Milky Way that is much less dense than the bulge \citep{bovy:2017b} -- then it follows that encounters are more common closer to the Galactic center.

One goal of our work was to estimate the stellar encounter rate as a function of observable quantities. We investigate the limits of what might be acheivable in Figure~\ref{fig:encounters_grid}, where each axis represents a quantity that might be derived  by exploiting the Gaia satellite's end-of-mission (likley by 2025) accuracy in parallax and proper motion measurement. Figure~\ref{fig:E_J_err} shows the expected percent error on both energy and angular momentum for stars in our sample, calculated using 10\% parallax errors, 20 km/s radial velocity error, and proper motion errors predicted by Gaia for 20 mag stars (chosen because G stars at most of the distances in our sample would be fainter than 19 mag). Errors on $E$ are low (many $<20$\%) for stars with the most energetic orbits, so it should be possible to observationally distiguish which stars experience the most encounters according to the left panel of Figure~\ref{fig:encounters_grid}.


Our goal was to get a \textit{fiducial} estimate of how many stellar encounters happen in the Milky Way bulge. As a result, our findings hold true for simple, ideal scenarios and don't necessarily reflect the complex nature of stellar populations.

For example, we found that 90\% of stars can expect to have encounters within a Hubble time, but not all stars in the bulge survive on the Main Sequence for that long \citep{zakhozhay:2013}. Our work does not take into account multiple generations of star formation, and can therefore be limited to conclusions about GKM stars. 

We also didn't take binarity of bulge stars into account. Studies estimate that 30-45\% of disk FGK stars are members of binary systems \citep{gao:2014}, but the binary fraction in the bulge remains unknown. The assumptions inherent in our work -- that close stellar encounters can destabilize planet orbits over time -- might not hold for circumbinary systems, which can have more complicated dynamics than systems with just one host star. 

We deliberately limit the scope of this work to a semi-analytical estimate of encounter rates for stars in the Milky Way bulge. Recent work suggests that planets that experience gravitational interactions with other stars can be stripped from their host stars or have their orbits destabilized \citep{li:2019,elteren:2019,cai:2019}. Other work suggests that the planet formation process can be stunted or halted altogether by photoevaporation from high amounts of radiation or tidal disruption from close stellar encounters \citep{vincke:2018,winter:2018a,winter:2018b}. Determining the effects of a stellar encounter on a planetary disk requires knowledge of so many factors beyond the encounter distance that we focused on here: mass ratio of the stars involved, speed of the encounter, relative inclination of the disks, etc.

\citet{bhandare:2019} simulated hyperbolic stellar fly-bys by varying many of these parameters and provided a helpful catalogue of planetary disk characteristics immediately following the encounter. They show that for a stellar encounter at 1000 AU where the mass ration between the perturber and planet host is 1, all particles are still bound to their host star and their orbital eccentricities are largely unchanged immediately after the encounter. A similar encounter at 100 AU  can strip away nearly 60\% of the particles around the host star and the remaining particles have been pushed onto highly eccentric orbits.

These outcomes change for a more extreme mass ratio. The most extreme ratio explored by \citet{bhandare:2019} is $m_{12} = M_{\mathrm{pert}}/M_{\mathrm{host}} = 50$. In that case, an encounter at 1000 AU leaves all test particles bound to the host star, but their orbits become more eccentric. A similar encounter at 100 AU will strip away $\sim97\%$ of the particles in the disk and many of the remaining particles will move on high eccentric orbits. 

The planetary disks in \citet{bhandare:2019} only extend to 100 AU and therefore provide no direct information on consequences for the Oort Cloud. It is not unreasonable to speculate, however, that the prevalence of stellar encounters within 1000 AU in the Milky Way bulge would disrupt the formation of structures like the Oort Cloud, which sits at around 5000 AU from the Sun in our own Solar System.

In future work, we plan to determine how common bulge encounters of different mass ratios are so that we can apply the results from \citet{bhandare:2019} specifically to the bulge. We also plan to explore how small changes in eccentricity immediately after the fly-by can affect the stability of the disk several million years later. More thought will also need to be given to the relative speeds of the stars involved in encounters, as that determines the amount of time that a perturber star exerts gravitational influence on the host star system. Taking all of these into account will give us a much better understanding of the survival rate of planets in the Milky Way bulge. 


\section*{Acknowledgements}
M.A.S.M. is supported by the NSF Graduate
Research Fellowship under grant No. DGE 16-44869. KVJ's contributions were supported by
NSF grant AST-1715582. DMK's contributions were supported by the Alfred P. Sloan Foundation.



\bibliographystyle{mnras}
\bibliography{mybibliography} 


\bsp	
\label{lastpage}
\end{document}